\documentclass[aps,prl,twocolumn,amsmath,amssymb,superscriptaddress,showpacs,showkeys,preprintnumbers,nofootinbib]{revtex4-2}

\usepackage{dcolumn}
\usepackage{graphicx,color}
\usepackage{multirow}
\usepackage{textcomp}
\usepackage{physics}
\usepackage{dsfont}
\usepackage{relsize}
\usepackage{comment}

\usepackage{array, multirow, bigdelim, makecell, booktabs} 
\usepackage{tikz}	
\usetikzlibrary
{
	arrows,
	calc,
	through,
	shapes.misc,	
	shapes.arrows,
	chains,
	matrix,
	intersections,
	positioning,	
	scopes,
	mindmap,
	shadings,
	shapes,
	backgrounds,
	decorations.text,
	decorations.markings,
	decorations.pathmorphing,	
}

\begin{document}
\newcommand {\nc} {\newcommand}
  \nc {\eol} {\nonumber \\}
  \nc {\eoln}[1] {\label {#1} \\}
  \nc {\ve} [1] {\mbox{\boldmath $#1$}}
  \nc {\ves} [1] {\mbox{\boldmath ${\scriptstyle #1}$}}
  \nc {\mrm} [1] {\mathrm{#1}}
  \nc {\half} {\mbox{$\frac{1}{2}$}}
  \nc {\thal} {\mbox{$\frac{3}{2}$}}
  \nc {\fial} {\mbox{$\frac{5}{2}$}}
  \nc {\la} {\mbox{$\langle$}}
  \nc {\ra} {\mbox{$\rangle$}}
  \nc {\etal} {\emph{et al.}}
  \nc {\eq} [1] {(\ref{#1})}
  \nc {\Eq} [1] {Eq.~(\ref{#1})}

  \nc {\Refc} [2] {Refs.~\cite[#1]{#2}}
  \nc {\Sec} [1] {Sec.~\ref{#1}}
  \nc {\chap} [1] {Chapter~\ref{#1}}
  \nc {\anx} [1] {Appendix~\ref{#1}}
  \nc {\tbl} [1] {Table~\ref{#1}}
  \nc {\Fig} [1] {Fig.~\ref{#1}}
  \nc {\ex} [1] {$^{#1}$}
  \nc {\Sch} {Schr\"odinger }
  \nc {\flim} [2] {\mathop{\longrightarrow}\limits_{{#1}\rightarrow{#2}}}
  \nc {\textdegr}{$^{\circ}$}
  \nc {\inred} [1]{\textcolor{red}{#1}}
  \nc {\inblue} [1]{\textcolor{blue}{#1}}
  \nc {\IR} [1]{\textcolor{red}{#1}}
  \nc {\IB} [1]{\textcolor{blue}{#1}}
  \nc{\pderiv}[2]{\cfrac{\partial #1}{\partial #2}}
  \nc{\deriv}[2]{\cfrac{d#1}{d#2}}
\title{Impact of the  $^6$Li asymptotic normalization constant onto $\alpha$-induced reactions of astrophysical interest}
\author{C.~Hebborn}
\email{hebborn@frib.msu.edu}
\affiliation{Facility for Rare Isotope Beams, Michigan State University, East Lansing, Michigan 48824, USA}
\affiliation{Lawrence Livermore National Laboratory, P.O. Box 808, L-414, Livermore, California 94551, USA}
\author{M.~L.~Avila}
\affiliation{Physics Division, Argonne National Laboratory, Lemont, IL 60439, USA}
\author{K. Kravvaris}
\affiliation{Lawrence Livermore National Laboratory, P.O. Box 808, L-414, Livermore, California 94551, USA}
\author{G. Potel}
\affiliation{Lawrence Livermore National Laboratory, P.O. Box 808, L-414, Livermore, California 94551, USA}
\author{S. Quaglioni}
\affiliation{Lawrence Livermore National Laboratory, P.O. Box 808, L-414, Livermore, California 94551, USA}

\preprint{LLNL-JRNL-851130}

\date{\today}
\begin{abstract}
Indirect methods have become the predominant approach in experimental nuclear astrophysics for studying several low-energy nuclear reactions occurring in stars, as direct measurements of many of these relevant reactions are rendered infeasible due to their low reaction probability. Such indirect methods, however, require theoretical input that in turn can have significant poorly-quantified uncertainties, which can then be propagated to the reaction rates and  have a large effect on our quantitative understanding of stellar evolution and  nucleosynthesis processes.
We present two such examples involving $\alpha$-induced reactions,  $^{13}$C($\alpha,n)^{16}$O and $^{12}$C$(\alpha,\gamma)^{16}$O, {for which the low-energy cross sections} have been constrained with $(^6$Li$,d)$ transfer data.  In this Letter, we discuss how a 
first-principle calculation of $^6$Li leads to a 21\% reduction of  the $^{12}$C$(\alpha,\gamma)^{16}$O cross sections with respect to  a previous estimation. 
This calculation further resolves the discrepancy between recent measurements of the $^{13}$C$(\alpha,n)^{16}$O reaction and points to the need for improved theoretical formulations of nuclear reactions.   

\end{abstract}
\maketitle
%


\noindent 
 {\it Introduction:}   Nuclear fusion reactions involving the capture of  a helium-4 nucleus ($\alpha$-particle) from a light or medium-mass isotope to form a heavier nucleus and a neutron ($n$) or a high-energy photon ($\gamma$) are  {central to understanding} the  lifecycle of massive stars, from driving  {the nucleosynthetic processes that make them shine and evolve}~\cite{deBoerReview,Rauscher_2013,ARNOULD20031} to {determining their remnants after their eventual death}~\cite{Blackhole12C,Shen_2023,Fields_2016,RemnantSupernova12C}. For example, the $^{13}$C$(\alpha,n)^{16}$O and $^{22}$Ne$(\alpha,n)^{25}$Mg reactions are  the principal sources of neutrons fueling the slow neutron capture process (s-process) in asymptotic giant branch (AGB) stars,  which is responsible for the formation of half of the elements heavier than iron~\cite{Kappeler2011157}.  Similarly, the $^{12}$C$(\alpha,\gamma)^{16}$O reaction is not only a key process in the sequence of helium burning reactions that produces carbon and oxygen in red giant and supergiant stars~\cite{Quaglioni2015} but also  determines the ratio of {the amount} of $^{12}$C to that of $^{16}$O, which has profound repercussions on the later evolutionary phases and nucleosynthesis events of these stars, and their ultimate fate once they explode as supernovae~\cite{deBoerReview}.   Arriving at  a quantitative  and more fundamental understanding of the life and death of massive stars requires accurate and precise knowledge of $\alpha$-induced reaction rates at  stellar energies.  In the case of $^{12}$C$(\alpha,\gamma)^{16}$O, ideally the reaction rate would need to be known within $\approx 10\%$ uncertainty or less at center of mass energies of $\approx 300$ keV~\cite{BUCHMANN2006254,deBoerReview}.

Typically, $\alpha$-induced reactions are very difficult or impossible to measure in the range of (low) energies where they occur in stars because the Coulomb repulsion between the $\alpha$ particle and the nucleus suppresses the reaction probability (cross section) below the background of cosmic rays.   Underground facilities, such as the Gran Sasso National Laboratories (LUNA)~\cite{LUNA} or the China Jinping Underground Laboratory (JUNA)~\cite{JUNA}, help reduce this background   and reach the low energies relevant for astrophysics.  However, a direct measurement of, e.g., the $^{12}$C$(\alpha,\gamma)^{16}$O reaction below $\approx 300$ keV remains unfeasible and down extrapolations from higher-energy measurements rely on theory. 

For the description of low-energy reactions involving systems made of more than $A$=12 nucleons, where accurate microscopic predictions based on validated models of the nuclear interactions~\cite{Quaglioni2020,Navratil2020} are out of reach and few-body models using effective potentials between structureless reactants~\cite{IGO1958,LovellUQJPG,OpticalPotentialReview} do not provide the required predictive power, phenomenological R-matrix theory~\cite{PhysRev.72.29,RevModPhys.30.257} has been the tool of choice. In this technique, cross sections are reconstructed from a relatively small number of parameters that have a physical meaning and are adjusted to reproduce available experimental data.    R-matrix analyses  have been used extensively to evaluate S-factors\footnote{Astrophysical S-factors correspond to the cross sections rescaled to remove the effect of the Coulomb barrier.} of astrophysical interest, including $^{13}$C$(\alpha,n)^{16}$O~\cite{LUNA13Calpha,JUNA13Calpha} and $^{12}$C$(\alpha,\gamma)^{16}$O~\cite{deBoerReview}, and extrapolate them down to stellar energies.  However, the fit of the parameters can  result in sizeable uncertainties on the extrapolated S-factors, in particular when different data sets are inconsistent with each other. Moreover, the presence of loosely-bound states can further influence the extrapolation process~\cite{PhysRevC.59.3418}. To reduce  uncertainties, parameters related to the asymptotic normalization of bound-state wave functions can be fixed using information gleaned from measurements of $(^6$Li$,d)$ [or $(^7$Li$,t)$] $\alpha$-transfer processes,  in which  an $\alpha$-particle is transferred from a loosely-bound $^6$Li  ($^7$Li) to another nucleus $A$ to form the $A$-$\alpha$  state of interest~\cite{Brune12C6Lid,Melina12C6lid,AvilaThesis,ReviewHammache2020}. 

In this Letter, we  discuss how a recent
first-principle prediction for the $^6$Li nucleus~\cite{6LiHebborn} influences the low-energy properties extracted from $(^6$Li$,d)$ transfer data and  explains the discrepancy between  two recent {R-matrix} evaluations  of the $^{13}$C$(\alpha,n)^{16}$O reaction  by the LUNA~\cite{LUNA13Calpha} and JUNA collaborations~\cite{JUNA13Calpha}. We also estimate the significant impact of this $^6$Li prediction on the  R-matrix evaluation of the $^{12}$C$(\alpha,\gamma)^{16}$O rate, and argue for future theoretical and experimental studies to improve the evaluation of additional $\alpha$-induced reactions.

{\it  Peripheral reactions to extract ANCs:} 
Non-resonant {low-energy} radiative {$\alpha$-}capture reactions $A+\alpha\rightarrow B  +\gamma$  dominated by electric transitions are peripheral, i.e., do not probe the interior of the $A$-$\alpha$ bound state (Fig. \ref{fig_anc}) and their cross sections scale with the square of {its asymptotic normalization constant (ANC, denoted as
$\mathcal{C}_{A{\textrm -}\alpha}$)}~\cite{ANCmehtodRev}.  The  low-energy $\alpha$-capture cross section can then be accurately approximated as~\cite{PhysRevC.61.025801,ANCmehtodRev} 
\begin{equation}
    \sigma_{\alpha,\gamma} \simeq (\mathcal{C}_{A{\textrm -}\alpha})^2\frac{\hat{\sigma}_{\alpha,\gamma} }{b^2_{A{\textrm -}\alpha}}, \label{eq1}\,
\end{equation}
where $\hat{\sigma}_{\alpha,\gamma}$ and $b_{A{\textrm -}\alpha}$ are the cross section  and ANC obtained in a two-body model calculation that treats both the target nucleus ($A$) and the $\alpha$ as point particles ($b_{A{\textrm -}\alpha}$ is also referred as the single-particle ANC).  {More generally, the $(\mathcal{C}_{A{\textrm -}\alpha})$ ANC is connected with the $\alpha$-partial  width of the corresponding state in the $A+\alpha$ system~\cite{deBoerReview,ANCmehtodRev}. {It is thus also an essential ingredient in the description and
	analysis of many low energy processes, such as the $^{13}$C$(\alpha, n)^{16}$O reaction discussed in this work, but also in multiple other cases where $\alpha$ particles are involved in either the incoming or the outgoing channel~\cite{JAYATISSA2020135267,ANCmehtodRev}.}

Transfer reactions at energies around the Coulomb barrier  are  also peripheral,  and hence exhibit a similar proportionality with ANCs.  For example, the cross section for $A+^6$Li$\rightarrow B (\equiv A+\alpha)+d$ at low energies can be  accurately evaluated as
\begin{equation}
    \sigma_{^6{\rm Li},d}  \simeq (\mathcal{C}_{A{\textrm -}\alpha})^2(\mathcal{C}_{\alpha{\textrm -}d})^2\frac{\hat{\sigma}_{^6{\rm Li},d}}{b^2_{A{\textrm -}\alpha}b^2_{{\alpha}-d}}\,, \label{eq2}
\end{equation}
where, similar to before, $\hat{\sigma}_{^6{\rm Li},d}$  and $b_{\alpha{\textrm -}d}$ are the cross section and ANC obtained in simplified calculations that treat  $^6{\rm Li}$, $d$ and  $\alpha$ as point particles.  If the ANC  of the $^6$Li nucleus $\mathcal{C}_{\alpha{\textrm -}d}$  is well known, one can use {{Eq.~(\ref{eq2})} and} {experimental data on the transfer reaction to} accurately extract $\mathcal{C}_{A{\textrm -}\alpha}$ by rescaling the theoretical cross sections $\hat{\sigma}_{^6{\rm Li},d}$, typically evaluated within the distorted-wave Born approximation (DWBA),  to the data~\cite{ANCmehtodRev}. These calculations usually consider only the s-wave ANC of  $^6$Li and neglect the d-wave ANC, which is two order of magnitude smaller~\cite{GK99,6LiHebborn}.

\begin{figure}
    \centering
    \includegraphics[width=\linewidth]{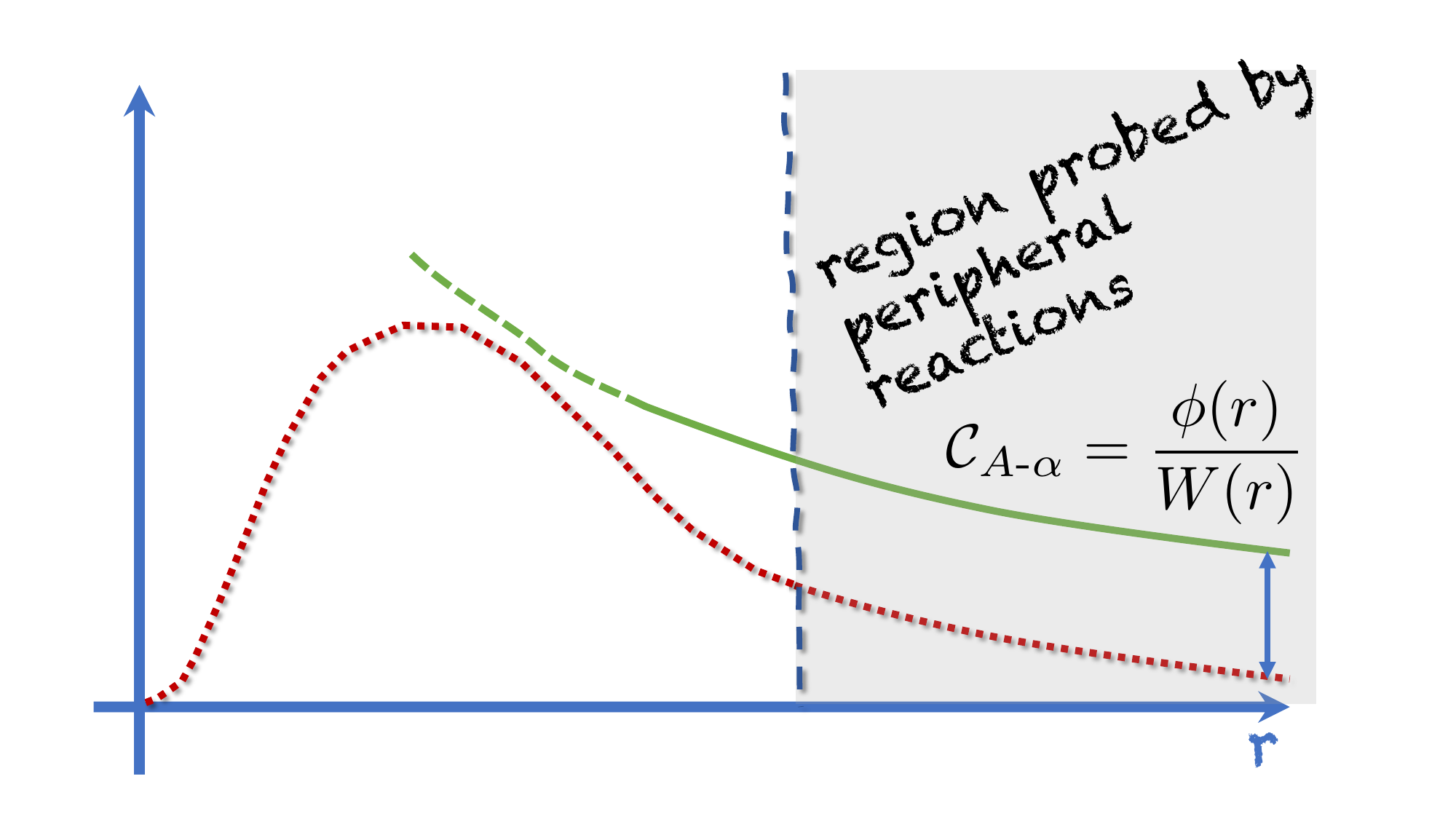}
    \caption{Schematic representation of the radial part of the $A{\textrm -}\alpha$ wavefunction $\phi(r)$ (dotted red line) and the Whittaker function $W(r)$ (green continuous line). They differ  by a constant factor, namely the asymptotic normalization constant $\mathcal C_{A{\textrm -}\alpha}$. Peripheral reactions are only sensitive to the asymptotic behavior of the $A{\textrm -}\alpha$ wavefunction (shaded area).}
    \label{fig_anc}
\end{figure}
A frequently adopted value of the $s$-wave $\alpha$-$d$ ANC in the analysis of peripheral $(^6$Li$,d)$ transfer reactions~\cite{Brune12C6Lid,Melina12C6lid} has been $(\mathcal{C}_{\alpha{\textrm -}d})^2=5.3 \pm 0.5$~fm$^{-1}$. {This value was} determined by Blokhintsev \etal~\cite{PhysRevC.48.2390} using three different methods: 1) by analytic continuation of the $\alpha$-$d$ scattering phase shifts, 2) by computing the (two-body) $\alpha$-$d$ bound state using an interaction fitted to reproduce these phase shifts, and 3) by solving the Faddeev equations for the (three-body) $\alpha$-$n$-$p$ system using phenomenological $\alpha$-nucleon interactions and neglecting the $\alpha$-$p$ Coulomb repulsion.  Unfortunately,  the uncertainties associated with the extrapolation procedure used in the first two methods and the ambiguity in the choice of the $\alpha$-nucleon interactions combined with the omission of the Coulomb potential in the third method  have not been quantified, raising the prospect for previously unrecognised systematic errors in the ANC determination. Other evaluations~\cite{NWS01,GK99} of $ \mathcal{C}_{\alpha{\textrm -}d}$ relying on  $\alpha$-$d$ phenomenological potentials   are consistent with the values provided by Blokhintsev \etal~but none of them have quantified the parametric uncertainties associated with the fit of the  $\alpha$-$d$ interaction, which can be sizeable~\cite{IGO1958,LovellUQJPG,OpticalPotentialReview}. 
 
A new accurate determination of $\mathcal{C}_{\alpha{\textrm -}d}$ through  (six-body) predictions of the $^6$Li {system} starting from two- and three-nucleon forces derived within chiral effective field theory, recently became available~\cite{6LiHebborn}. These calculations, obtained within the framework of the \textit{ab initio} no-core shell model with continuum (NCSMC)~\cite{NCSMCPhysScripta}, treat the $\alpha$-$d$ scattering and bound $^6$Li state on equal footing and accurately reproduce the low-energy properties of the system, including the $\alpha(d,\gamma)^6$Li capture rate and  $\alpha$-$d$ elastic scattering at energies below 3 MeV. Contrary to the determination of Blokhintsev \etal, the uncertainties of these microscopic calculations stem from two clearly identified sources: the choice of chiral Hamiltonian and the convergence with respect to the size of the NCSMC model space {(see Supplemental Material of Ref.~\cite{6LiHebborn})}. Both uncertainties  were significantly reduced by introducing a fine-tuning correction  to exactly reproduce the experimental binding energy of the $^6$Li ground state.  Compared to  this first-principle prediction, $(\mathcal{C}_{\alpha{\textrm -}d})^2=6.864 \pm 0.210$~fm$^{-1}$, the ANC of Blokhintsev \etal~\cite{PhysRevC.48.1420} is 22\% smaller and exhibits larger uncertainties. 

{\it Impact  on s-process nucleosynthesis:} 
Because of its key role in  s-process nucleosynthesis, the $^{13}$C$(\alpha,n)^{16}$O reaction has been measured in underground facilities in the last two years\footnote{We cite here the year of the publications of the Phys. Rev. Lett. in which the data and analysis were published.}  at low energies relevant for astrophysics, i.e., at 0.23–0.3~MeV by the LUNA and at 0.24-1.9 MeV by the JUNA collaborations~\cite{LUNA13Calpha,JUNA13Calpha}. Although  the two data sets are consistent, the  R-matrix analyses used to extrapolate the data to lower energies lead to different S-factors. The JUNA collaboration suggested that this discrepancy is explained by the use of a different {$\alpha$-ANC of the } $1/2^+$  state in $^{17}$O, {located 3 keV below the $\alpha$-threshold,} which   determines
the overall normalization of the S-factor  at
the energies of astrophysics interest~\cite{ANCmehtodRev,JUNA13Calpha,LUNA13Calpha}.  In the work of Ref.~\cite{JUNA13Calpha} this ANC is treated as a floating parameter in the R-matrix analysis of the data, while in the analysis of Ref.~\cite{LUNA13Calpha} it was fixed to the value inferred by Avila~\etal~from a sub-Coulomb $(^6$Li$,d)$ transfer experiment~\cite{PhysRevC.91.048801} (Fig.~\ref{Fig17O}). In turn, however, the work of Ref.~\cite{PhysRevC.91.048801} had  adopted the s-wave $\alpha$-$d$ ANC of Blokhintsev \etal~{to extract the ANC of the $1/2^+$ state in $^{17}$O following the relationship described in Eq.~\eqref{eq2}.}

By {reinterpreting} the transfer data of Ref.~\cite{PhysRevC.91.048801} with the more accurate ANC obtained in Ref.~\cite{6LiHebborn}, we extract\footnote{The so-called Coulomb-modified ANC $\tilde{\mathcal{C}}$ is directly proportional to the usual ANC and given by $\tilde{\mathcal{C}} = \mathcal{C}\times \ell !/\Gamma(\ell+1+{|\eta|})$, with $\ell$ being the relative angular momentum between the fragments and $\eta$ the Sommerfeld parameter~\cite{ANCmehtodRev}.}  $(\tilde {\mathcal{C}}^{1/2+}_{\alpha{\textrm -}^{13}\rm C})^2=2.8 \pm 0.5$~fm$^{-1}$, consistent with the value obtained by the JUNA collaboration, thus explaining the difference between the two R-matrix analyses. {The new calculated value of $(\mathcal{C}_{\alpha{\textrm -}d})^2$ contributes by just 0.086 fm$^{-1}$ to the quoted uncertainty. The rest is associated with the experimental and theoretical  uncertainties arising from the ($^{6}$Li,$d$) experiment used to extract the $\alpha$-$^{13}$C ANC,  as described in \cite{PhysRevC.91.048801}. This is in sharp contrast with the previous analysis of \cite{PhysRevC.91.048801}, where the Blokhintsev $(\mathcal{C}_{\alpha{\textrm -}d})^2$ value contributed to the total uncertainty by 0.34 fm$^{-1}$, a much more significant amount.}  The re-evaluated ANC from the $(^6$Li$,d)$ transfer data also agrees with  the values extracted from  $(^{11}$B$,^7$Li$)$~\cite{11B7Li12Cdata} and $(^7$Li$,t)$~\cite{PhysRevC.77.042801} transfer data
(Fig.~\ref{Fig17O}).

\begin{figure}
    \centering
    \includegraphics[width=\linewidth]{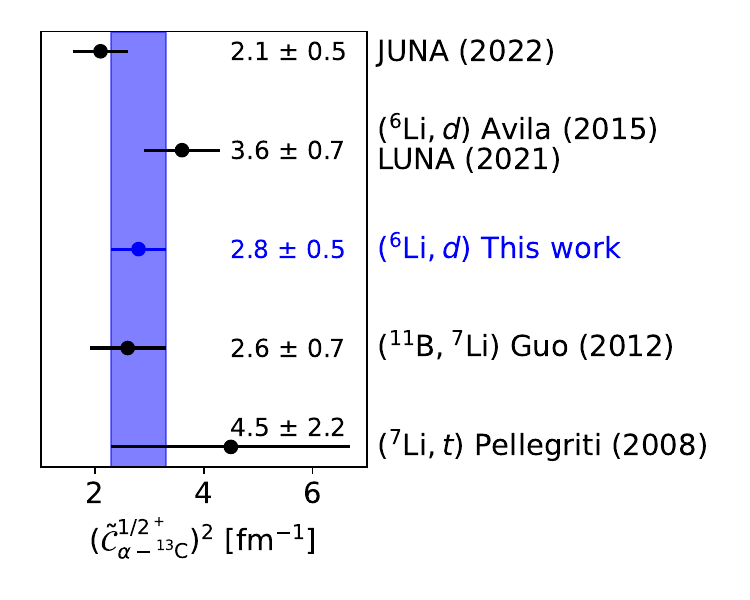}
    \caption{Comparison of  {the} square of {the} Coulomb-modified ANC of {the} $1/2^+$ threshold state in $^{17}$O, $(\tilde {\mathcal{C}}^{1/2+}_{\alpha{\textrm -}^{13}\rm C})^2$,  deduced {from:} {the} R-matrix analysis of the JUNA collaboration~\cite{JUNA13Calpha}, the  values extracted from $(^6$Li$,d)$ data~\cite{PhysRevC.91.048801} using the $(\mathcal{C}_{\alpha{\textrm -}d})^2$ of Ref.~\cite{PhysRevC.48.1420}, our {reinterpretation} of the $(^6$Li$,d)$ data~\cite{PhysRevC.91.048801} using a  first-principle prediction of $(\mathcal{C}_{\alpha{\textrm -}d})^2$~\cite{6LiHebborn}, the values inferred from $(^{11}$B$,^7$Li$)$  data~\cite{11B7Li12Cdata,SHEN2019134820}, $(^{7}$Li$,t)$  data~\cite{PhysRevC.77.042801}. The numbers in the figure correspond to the values of {the corresponding} ANCs and their uncertainties.}
    \label{Fig17O}
\end{figure}
This work  supports the JUNA  analysis~\cite{JUNA13Calpha} of the thermonuclear reaction rate~\cite{NACRE13C,REACLIB13C}, which is smaller than previous evaluations. This decrease in the flux of neutrons  impacts the abundances of s-process branching point and heavier elements, in particular $^{60}$Fe, $^{152}$Gd and $^{205}$Pb~\cite{LUNA13Calpha}. 

Interestingly, the rate of the other major neutron source in the s-process, $^{22}$Ne$(\alpha,n)^{25}$Mg, has also been constrained using the $\alpha$-partial width of resonances of $^{26}$Mg  extracted from  $(^6$Li$,d)$ data using the the s-wave $\alpha$-$d$ ANC of Blokhintsev \etal~\cite{JAYATISSA2020135267}.  Based on the larger value of the $\alpha$-$d$ ANC of Ref.~\cite{6LiHebborn}, we expect that this  $^{22}$Ne$(\alpha,n)^{25}$Mg rate is overestimated, and that the overall neutron flux  in the s-process is even smaller than what is currently evaluated. Because the  $^{22}$Ne$(^6$Li$,d)^{26}$Mg transfer reaction populates unbound $^{26}$Mg   states, the DWBA analysis of the data should be revisited to account for model uncertainties associated with the approximation introduced by treating these states as bound and then extrapolating their properties up to energies in the continuum~\cite{PhysRevC.90.042801}. We reserve this study for future work, as some development in reaction theory is first needed to adequately describe  transfer reactions to states in the energy continuum, e.g., through the generalization of the Green's function based formalisms~\cite{GFT,GFTapp,GFK} to $\alpha$-transfer reactions.

{\it Impact on $^{12}{\rm C}(\alpha,\gamma)^{16}{\rm O}$:} {A recent} R-matrix analysis {of $^{16}$O data} (including $^{12}$C$(\alpha,\gamma)^{16}$O capture,  $\beta$-delayed $\alpha$-emission,  $\alpha$-$^{12}$C elastic scattering measurements, and recent evaluations of bound- and resonant-state properties) found that the extrapolated $^{12}$C$(\alpha,\gamma)^{16}$O S-factor at stellar energies of $\approx 300$~keV depends strongly on the $\mathcal{C}_{\alpha{\textrm -}^{12}{\rm C}}$ ANC of the $1^-$ and $2^+$  loosely-bound states in $^{16}$O~\cite{deBoerReview}.
\begin{table}[t]
	\centering
	\begin{tabular}{cccccccccl}
		\hline\hline\\[-2mm]
		&&  && & \multicolumn{4}{c}{$(\mathcal {C}_{\alpha{\textrm -}^{12}\rm C}^{J^\pi})^2$ }\\[1mm]
		\cline{7-9}\\[-2mm]
		$J^\pi$ && $E_{\rm ex}$ && Probe &&  Past Work && This Work \\[1mm]
		\hline\\[-2mm]
		$0^+$&& 6.05 &&$(^6$Li$,d)$~\cite{Melina12C6lid}&& 
		2.43(30) && 1.88(16)& \hspace{-0.75em}\rdelim\}{1}{*}[${}\times 10^{6}$ ]\\[1mm] 
		\\
		$3^-$&& 6.13&&$(^6$Li$,d)$~\cite{Melina12C6lid}&& 1.93(25)&& 1.49(14)& \hspace{-0.75em}\rdelim\}{1}{*}[${}\times 10^{4}$ ]\\[1mm]
		\\
		\multirow{5}{*}{$2^+$}&& \multirow{5}{*}{6.92} &&$(^6$Li$,d)$~\cite{Brune12C6Lid}&& 1.24(24) && 0.96(16) & \hspace{-0.75em}\rdelim\}{5}{*}[${}\times 10^{10}$ ]\\[1mm]  
		&&  &&$(^6$Li$,d)$~\cite{Melina12C6lid}&& 1.48(16) && 1.14(7) \\[1mm]
		&&  &&$(^7$Li$,t)$~\cite{Brune12C6Lid}&& 1.33(29)&&\\[1mm]
		&&  &&$(^7$Li$,t)$~\cite{Oulebsir}&& 2.07(80) &&\\[1mm]
		\\
		\multirow{3}{*}{$1^-$} &&\multirow{3}{*}{7.12}&&$(^6$Li$,d)$~\cite{Brune12C6Lid} &&4.33(84)  &&  3.34(58)  & \hspace{-0.75em}\rdelim\}{3}{*}[${}\times 10^{28}$ ]\\ 
		&&  &&$(^6$Li$,d)$~\cite{Melina12C6lid}&&4.39(59) &&3.39(34)   \\[1mm]
		&&  &&$(^7$Li$,t)$~\cite{Oulebsir}&&4.00(138)  &&& \\[1mm]
		\hline\hline
	\end{tabular}
	\caption{Square of ANCs (in units of fm$^{-1}$) for the $0^+$, $3^-$, $2^+$ and $1^-$ bound  states of $^{16}$O extracted from $^{12}$C$(^6$Li$,d)^{16}$O~\cite{Brune12C6Lid,Melina12C6lid} and $^{12}$C$(^7$Li$,t)^{16}$O~\cite{Oulebsir,Brune12C6Lid} data. {The first two columns indicate the excitation energy (in MeV), spin, and parity, of the corresponding state, and the third column specifies the reaction used for the extraction of the 
			ANC}. {In the last two columns, we compare}  the original evaluation of the ANC  using  $(\mathcal{C}_{\alpha{\textrm -}d})^2$  of Ref.~\cite{PhysRevC.48.2390} {with} the analyses using a first-principle prediction of  $(\mathcal{C}_{\alpha{\textrm -}d})^2$~\cite{6LiHebborn}.}
	\label{Tab1}
\end{table}
To arrive at their best R-matrix fit, the authors adopted (as fixed parameters) the ANCs of these and the other $^{16}$O bound  states determined from $(^6$Li$,d)$ reactions~\cite{Brune12C6Lid,Melina12C6lid} using the $\mathcal{C}_{\alpha-d}$ value of Blokhintsev \etal~({fourth} column of Table~\ref{Tab1}). 
As in the previous case, the new accurate prediction of $\mathcal{C}_{\alpha{\rm -}d}$ from Ref.~\cite{6LiHebborn} impacts the extracted $^{16}$O $(\mathcal {C}_{\alpha{\textrm -}^{12}\rm C}^{J^\pi})^2$ values, yielding ANCs that are $\approx 22\%$ smaller  and with reduced uncertainties (fifth  column of Table~\ref{Tab1}). 
These new evaluations are consistent with the $(\mathcal{C}^{1^-}_{\alpha{\textrm-}^{12}{\rm C}})^2$ and $(\mathcal{C}^{2^+}_{\alpha{\textrm-}^{12}{\rm C}})^2$ values extracted from $(^7$Li$,t)$ transfer reactions in Ref.~\cite{Oulebsir} and \cite{Brune12C6Lid} respectively, but are  in tension with the $2^+$ ANC from Ref.~\cite{Oulebsir} (Table~\ref{Tab1}). The $^7$Li ANC  used in these DWBA analyses was determined using similar approaches as in Ref.~\cite{PhysRevC.48.2390} and model uncertainties are similarly not fully quantified~\cite{KAJINO1988475,PhysRevC.48.1420}. In this respect, a first-principle prediction of the $^7$Li ANC $\mathcal{C}_{\alpha-t}$ would be desirable and may resolve the tension  between the $\mathcal{C}_{\alpha{\textrm -}^{12}{\rm C}}$ values extracted from $(^6$Li$,d)$ and $(^7$Li$,t)$ transfer data. 

To illustrate the impact of the $\mathcal {C}_{\alpha{\textrm -}^{12}\rm C}^{J^\pi}$ values determined in this work on the low-energy $^{12}$C$(\alpha,\gamma)^{16}$O S-factor, we perform a reduced R-matrix calculation\footnote{We also included the cascade transitions, resulting from  $\gamma$-ray de-excitations from the excited states to the ground state of $^{16}$O.} starting from the best fit of Ref.~\cite{deBoerReview}, keeping only the  $^{16}$O states with excitation energies up to 10.36 MeV.  At low energies, the S-factor from such reduced R-matrix calculation is close to the comprehensive fit of Ref.~\cite{deBoerReview} (Fig.~\ref{Fig1}). {We then rescale the ANCs of the bound states and perform an R-matrix calculation without refitting the rest of the parameters.} As expected for a peripheral reaction, upon changing the $\alpha$-$^{12}$C ANCs to the values determined in this work we obtain a reduced  S-factor.  At 300 keV, the S-factor is almost exactly proportional to $(\mathcal {C}_{\alpha{\textrm -}^{12}\rm C}^{J^\pi})^2$ and  is reduced by 21\% with respect to the original fit. {It is worth noting 
that none of the 
S-factors are 
consistent with both 
data sets~\cite{PhysRevC.86.015805,Schurmann05}. Because of the 
need of renormalizing
measurements, and
the difficulties in 
extracting the overall
normalization, 
R-matrix analyses
lack predictive power
for the determination of ANCs. It also indicates that these data alone do not sufficiently constrain  the S-factor at stellar energies.}

\begin{figure}
    \centering
    \includegraphics[width=\linewidth]{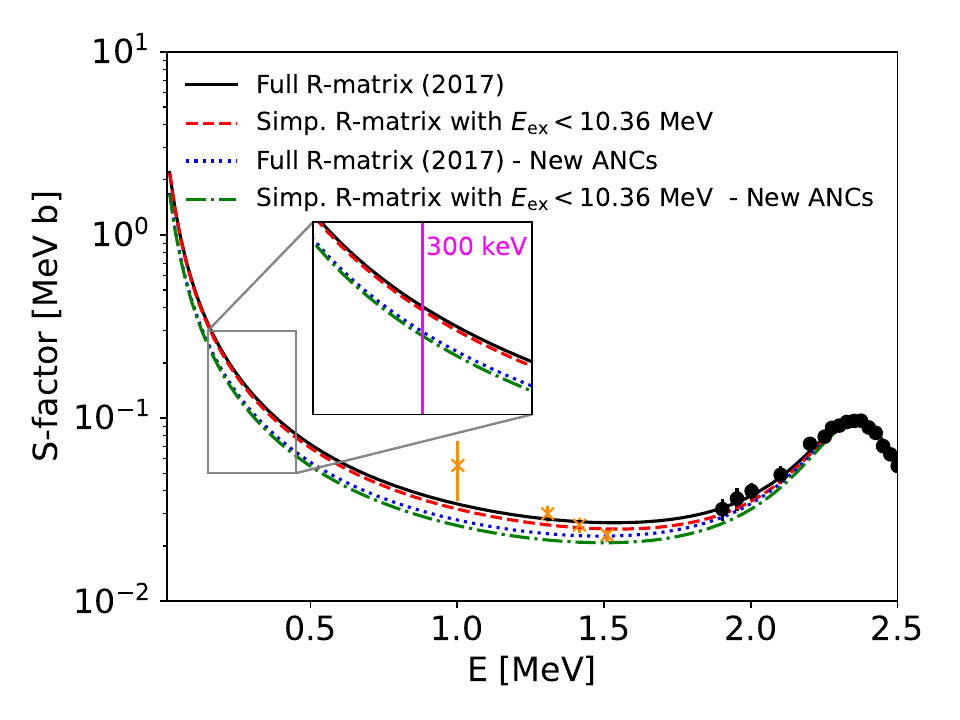}
 
    \caption{Comparison of the $^{12}$C$(\alpha,\gamma)^{16}$O S-factors as a function of the relative $^{12}$C-$\alpha$ relative energy, obtained with the R-matrix analysis of deBoer \etal~\cite{deBoerReview} (solid black line), a simplified version in which only the low-energy states and the $\alpha$-channel are included (dashed red line), and predictions using our $(\mathcal {C}_{\alpha{\textrm -}^{12}\rm C}^{J^\pi})^2$ displayed in Table~\ref{Tab1} (dotted blue and dash-dotted green lines).  These S-factors have been obtained using the code {\sc AZURE2}~\cite{AZURE2_1,AZURE2_2}, and  the input files are provided in Supplemental Material~\cite{SupplMat}. The total S-factor data from Refs.~\cite{Schurmann05} (black circles) and~\cite{PhysRevC.86.015805} (orange crosses) are also shown for comparison. The inset show the S-factors around the stellar energies (in magenta).
}
    \label{Fig1}
\end{figure}

 The $21\%$ reduction of the S-factor at stellar energies will increase the ratio of the $^{12}$C to $^{16}$O abundances. Quantifying the impact of our new evaluations of the $\mathcal{C}_{\alpha{\textrm -}^{12}{\rm C}}$ ANCs on the other carbon burning scenarios at temperature $1$-$10$~GK would require reevaluating the S-factor up to 6~MeV by means of a more complete R-matrix fit, similar to the one presented in Ref.~\cite{deBoerReview}, which included multiple reactions channels and data sets and used a robust statistical framework to quantify the uncertainties.

{\it Conclusions and prospects:} We have demonstrated how a recent first-principle prediction for $\mathcal{C}_{\alpha{\textrm-}d}$  {impacts} S-factors of astrophysical interest, for both the s-process and helium burning nucleosynthesis. Using this new $\mathcal{C}_{\alpha{\textrm -}d}$,  the values for the $^{13}$C-$\alpha$ ANC extracted from $(^{6}$Li$,d) $ data now agree with values extracted from other $\alpha$-transfer probes. Our analysis further provides an explanation  for the discrepancy between the two recent LUNA and JUNA evaluations of the $^{13}$C$(\alpha,n)^{16}$O S-factors and reaction rates, favoring the recent JUNA evaluation~\cite{JUNA13Calpha}. Since  the other principal neutron source in the s-process, i.e., $^{22}$Ne$(\alpha,n)^{25}$Mg, was also evaluated using $(^{6}$Li$,d) $  data, our analysis suggests that the neutron flux in the s-process is  smaller than what is currently evaluated.

The first-principle prediction of  $\mathcal{C}_{\alpha{\textrm -}d}$ also impacts the  $^{12}$C$(\alpha,\gamma)^{16}$O reaction rate. We find that the S-factor at 300 keV is reduced by 21\%, compared to Ref.~\cite{deBoerReview}. This reduction, when propagated into a nucleosynthesis reaction network, will  increase the abundance ratio of $^{12}$C to $^{16}$O and  impact the abundances of heavier elements.
This work advocates for a new R-matrix analysis, constrained with our new evaluation of the $\mathcal{C}_{^{12}{\rm C}{\textrm -}\alpha}$,  of the $^{12}$C$(\alpha,\gamma)^{16}$O reaction rate over the all range of energies relevant for astrophysics. 

{Our approach will have a general impact on  the accuracy and reliability of indirect measurements of reaction rates of astrophysical interest by significantly reducing the uncertainty associated with $\mathcal{C}_{\alpha{\textrm -}d}$. The uncertainties in the extraction of the relevant partial widths are now dominated by the experimental errors and the reaction model used to describe the process. Within this context,} we have also pointed out some developments in reaction theory that should be conducted to improve our knowledge of any $\alpha$-induced reaction rate of astrophysical interest. Because  inaccuracies in the reaction model   used to extract structure information  from transfer data  propagate  to the astrophysical S-factor, it is crucial to push first-principle calculations, for which uncertainties {are} more straightforwardly quantifiable,  to heavier systems~\cite{KravvarisAlphaAlpha}.  In parallel, analysis of transfer data using few-body models could be improved to further constrain the S-factors:   these methods should be generalized to the transfer to unbound states, without requiring any extrapolation techniques, and the treatment of the reaction dynamics formalism should be improved, i.e., by going  beyond the one-step DWBA description. 
Finally, since  uncertainties can be reduced by comparing the structure information extracted from various transfer probes,  first-principle predictions of $^7$Li and $^{11}$B nuclei should be conducted and  compared with experiments that provide stringent constraint on  these ANCs, e.g., low-energy $d(^7$Li$,t)^6$Li measurements.\\
  
 \begin{acknowledgements}
\textit{Acknowledgments. } 
{C. H.\ would like to thank B. P.~Kay for inviting her for a  visit at Argonne National Laboratory, during which this work started.} 
This material is based upon work supported by the U.S. Department of Energy, Office of Science, Office of Nuclear Physics, under 
the FRIB Theory Alliance award no.\ DE-SC0013617, and under the Work Proposal no.\ SCW0498 (S.Q., K.K.). 
Prepared in part by LLNL under Contract No. DE-AC52-07NA27344. M.A.\ acknowledges the support of the U.S. Department of Energy, Office of Science, Office of Nuclear Physics, under contract number DE-AC02-06CH11357. Computing support for this work came from the
LLNL institutional Computing Grand Challenge program. 
\end{acknowledgements}

\bibliographystyle{apsrev}

\end{document}